\documentclass[pdflatex,sn-mathphys-num]{sn-jnl}

\usepackage{graphicx}%
\usepackage{multirow}%
\usepackage{amsmath,amssymb,amsfonts}%
\usepackage{amsthm}%
\usepackage{mathrsfs}%
\usepackage[title]{appendix}%
\usepackage{xcolor}%
\usepackage{textcomp}%
\usepackage{manyfoot}%
\usepackage{booktabs}%
\usepackage{algorithm}%
\usepackage{algorithmicx}%
\usepackage{algpseudocode}%
\usepackage{listings}%
\usepackage{booktabs}%
\usepackage[T1]{fontenc}
\usepackage{physics}

\theoremstyle{thmstyleone}%
\theoremstyle{thmstyletwo}%

\theoremstyle{thmstylethree}%
\newcommand{\gks}[1]{\textcolor{black}{ #1}}
\newcommand{\ayan}[1]{\textcolor{black}{ #1}}
\newcommand{\AF}[1]{\textcolor{black}{ #1}}

\begin{document}

\title[Article Title]{Free-space multi-user quantum network with high key rate}


\author*[1,2]{\fnm{Ayan Kumar} \sur{Nai}}\email{ayankrnai@gmail.com}

\author[1]{\fnm{G. K.} \sur{Samanta}}\email{gsamanta@pr.res.in}

\affil[1]{\orgdiv{Photonic Sciences Lab.}, \orgname{Physical Research Laboratory}, \orgaddress{\street{Navrangpura}, \city{Ahmedabad}, \postcode{380009}, \state{Gujarat}, \country{India}}}

\affil[2]{ \orgname{Indian Institute of Technology Gandhinagar}, \orgaddress{\city{Palaj, Gandhinagar}, \postcode{382055}, \state{Gujarat}, \country{India}}}



\abstract{
Emergent quantum networks are the essential ingredient for securely connecting multiple users worldwide, extensively deployed in both fibre and free-space. An essential element is the multiplexing of entanglement to multiple users, overcoming the peer-to-peer restriction of quantum key distribution (QKD), so far successfully shown in fibre-based architectures. Here, we demonstrate a free-space quantum space division multiplexing architecture using just one entanglement source to realise a fully connected twelve-channel quantum network for seamless QKD connections between six users.  The network achieves record coincidence rates exceeding $3 \times 10^{4}$ s$^{-1}$ between any pair of nodes on the network, for sifted key rate of over 400 kbps. Our approach overcomes the active switching hurdle that has hindered the free-space deployment of quantum multiplexing, is fully passive, easily scalable to more nodes and compatible with fibre-based integration, thus opening a new path to scalable and resource-efficient quantum networks that utilise free-space links.

}




\maketitle

\section{Introduction}\label{sec1}

Quantum key distribution (QKD), offering unconditional security based on the principles of quantum mechanics, has evolved as the mature and practical application of quantum information science \cite{Nielsen10}. Over the past two decades, it has successfully transitioned from theoretical proposals \cite{Bannett14, Ekert91} to field-deployable systems, especially between two users, with demonstrations in lab setup \cite{Bennett1992, Jennewein2000}, over fiber-optic networks \cite{korzh2015, yin2016, wengerowsky2020}, free-space links \cite{ursin2007, scheidl2009, Cai24}, and long distance satellite-based channels \cite{Liao2017, Bedington2017, Li2025} with the aim to provide secure key sharing for critical sectors such as defense, finance, and government communications. However, realizing the full potential of QKD and enabling its simultaneous connectivity among a large number of users worldwide necessitates the development of global quantum communication networks \cite{simon2017} analogous to the current classical communication infrastructure.

Recent advances have led to the demonstration of several types of quantum network architectures based on trusted node, quantum repeater, higher dimension entangled state, beam splitters, and spatial and frequency multiplexing \cite{Sasaki11, chen2023, Meter13, Avis2023b, Azum23, Mannalath23, Avis23a, townsend1997, liu2022, wengerowsky18, joshi2020, ortega2021} using different degrees of freedom, including polarization, frequency, and time. However, each architecture offers unique advantages and faces inherent limitations. For instance, while quantum memory-based networks \cite{Meter13, Avis2023b, Azum23} hold promise for universal applicability across network architectures, their practical realization remains challenging due to the stringent requirements of high-fidelity and long-coherence-time quantum memories \cite{Lei23}. Conversely, one-to-many user networks represent a simpler implementation but suffer from reduced bit rates and limited continuous connectivity owing to the use of multiple beam splitters or optical switching \cite{townsend1997, liu2022}. Trusted-node and high-dimensional entanglement-based networks, although conceptually powerful, are resource-intensive, posing constraints on scalability and widespread deployment \cite{Mannalath23, Avis23a}.

In contrast, wavelength-division multiplexing (WDM), a mature and widely used technique in classical networks \cite{Agrell2024}, has been extended to quantum sources at 1550 nm \cite{wengerowsky18, joshi2020, Liu2020}. Nevertheless, increasing the number of users in a quantum network by adding more frequency channels is fundamentally constrained by the limited photon flux or bit rate. Unlike classical communication systems, where bit rate can be enhanced by simply increasing the optical power, such an approach in quantum networks can violate the fundamental operating requirements of quantum light sources. Furthermore, channel dispersion and hardware-induced crosstalk must be compensated to preserve entanglement fidelity over long distances.

Recently, space-division multiplexing (SDM) \cite{Richardson2013, Puttnam21}, utilizing few-mode, multi-mode, and multi-core fiber architectures, has emerged as a complementary strategy to WDM in classical communication systems. These techniques harness the spatial degrees-of-freedom of light to significantly enhance channel capacity and are now being adapted for quantum communication \cite{liu2022, ortega2021, joshi2020}. Initial demonstrations of SDM-based quantum networks \cite{xavier2020, ortega2021}, as well as hybrid architectures combining WDM with beam-splitter-based SDM \cite{joshi2020, liu2022} using fiber channels at 1550 nm, have shown considerable promise as a scalable pathway toward high-capacity quantum networks. In a fiber-based WDM quantum network connecting $N$ users via a complete graph, $N(N-1)/2$ unique QKD links are required. This setup demands $N(N-1)/2$ distinct wavelength pairs from a single entangled photon source \cite{wengerowsky18}. A more resource-efficient architecture combining WDM and passive beam splitters has been proposed, reducing the requirement to only $N$ wavelength pairs while maintaining a fully connected $N$-user network \cite{joshi2020}. Most notably, the integration of both WDM and SDM has enabled a 40-user fully connected network comprising 780 QKD links \cite{liu2022}. \gks{The progress in fiber-based multi-user quantum networks has so far yielded relatively modest performance in terms of bit rate. Reported total coincidence rates are as low as 0.3 kbps in a laboratory setup using six sources with APD detection \cite{wengerowsky18}, with only a slight improvement to 0.56 kbps achieved over a 10 m fiber link using fifteen sources and SNSPD detection \cite{Liu2020}. Subsequent work reported a secure key rate of 4.1 kbps for a network formed with eight sources over channel lengths ranging from 10 m to 16 km using SNSPD detection \cite{joshi2020}. More recently, the secure key rate increased to 21 kbps for a network employing fifteen sources over 1–2 km patch-cord fiber links with SNSPD detection \cite{liu2022}.} \AF{As yet, the extension to free-space multiplexed links has remained largely unexplored due to the stringent resource requirements and the intrinsic challenges, such as beam divergence, pointing and tracking errors, and channel crosstalk, with satellite-based free-space QKD links remaining peer-to-peer.} Therefore, it is essential to explore multi-user networks for free-space links to establish a global quantum network while increasing the per-user bit rates for practical applications. 

Here, we present a novel spatial-sectioning and beam-splitter–based multiplexing architecture that enables a fully connected, free-space multi-user quantum communication network with a total sifted key rate of 407 kbps and a secure key rate
of 76 kbps across all user pairs. For any even number of users ($N>3$), the scheme requires only $N/2$ entangled-photon sources yet supports $(N(N-2)/2)$ unique user pairs, offering a highly scalable platform for multi-user entanglement distribution. By harnessing the intrinsic spatial and temporal correlations and quantum randomness \ayan{\cite{nai2025beam, nai2025device}} of the spontaneous parametric down-conversion (SPDC) process, \AF{we section the annular emission ring of an entangled-photon source and multiplex the sections with passive optical elements.} In a proof-of-concept demonstration, we establish a 12-channel QKD network among $N=6$ users, realized with only $N/2=3$ spatially sectioned entangled-photon sources derived from a single source. We achieved high performance across all user pairs, with coincidence count rates exceeding 30 kHz, entanglement visibility above 80$\%$, and average quantum state fidelity of 90$\%$. The proposed architecture is inherently scalable, as the number of entangled photon sources can be increased without compromising performance metrics. This is achieved by simultaneously increasing the pump laser power and exploiting the enlarged annular ring profile from a single SPDC source. Additionally, the integration of WDM and/or one-to-many beam splitters enables a significant expansion in the number of connected users without necessitating additional entangled photon sources or the use of central trusted nodes. The demonstrated architecture offers a practical route toward high-rate, globally deployable quantum communication networks, particularly suited for satellite-based free-space links. 

\section{Experiment}

\begin{figure}[H]
    \centering
    \includegraphics[width=\linewidth]{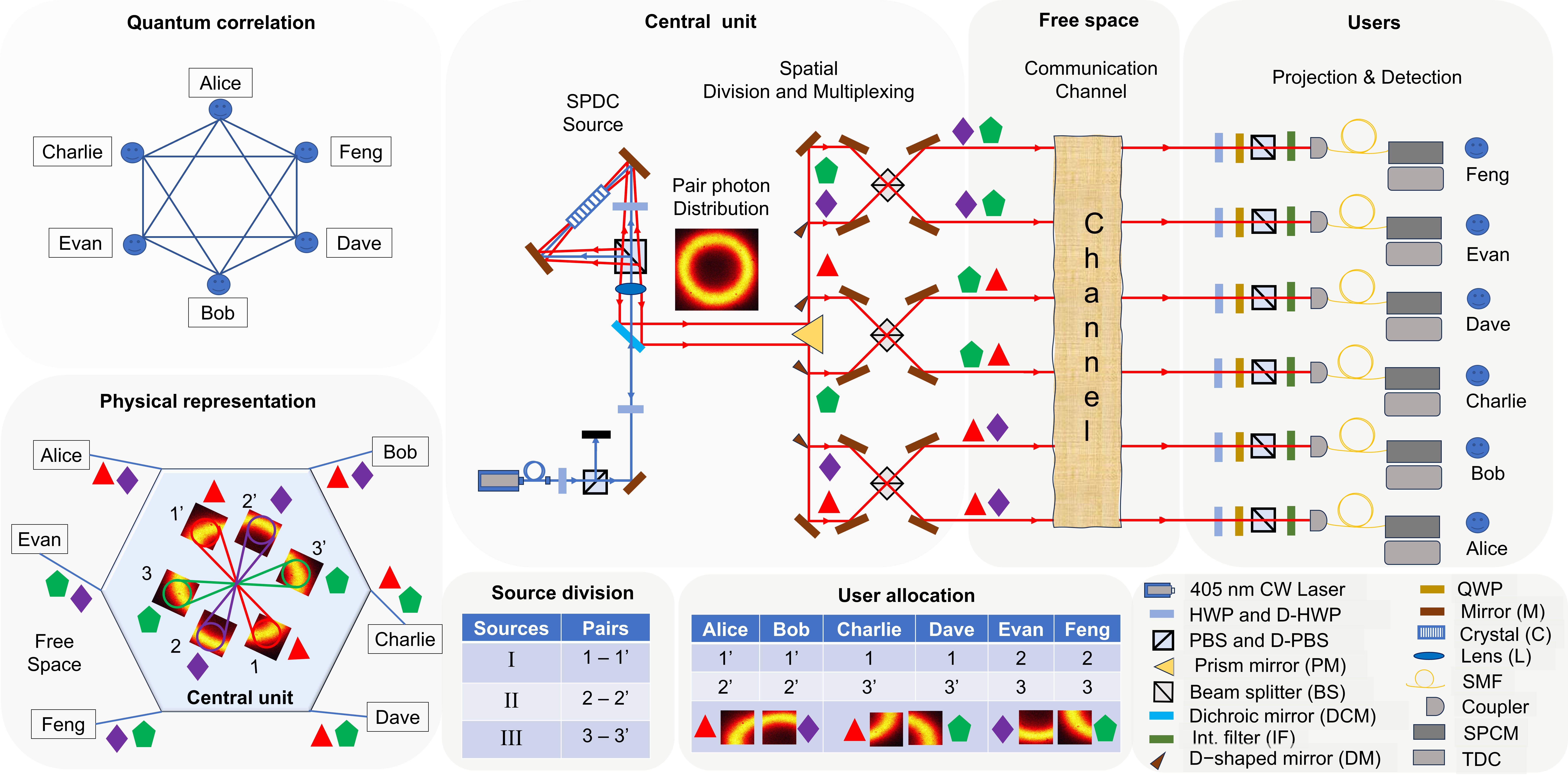}
    \caption{\textbf{Conceptual scheme and experimental setup of the multi-user quantum network architecture.} On the left, we have illustrated the concept of using two layers of abstraction. The top layer shows the fully connected mesh of 12 channels connecting six users through quantum correlation. The bottom layer shows the physical representation of the network topology, with the implementation shown on the right. The topology of the quantum network is established through the central unit, which derives three entangled photon sources from a single source and distributes them between six users through multiplexing.  The complete end-to-end implementation has three sections: central unit, free space channel, and users. Again, the central unit has two subsections: SPDC source and spatial division and multiplexing. The SPDC source is realized using a single-frequency continuous-wave diode laser at 405 nm, with output power controlled via a half-wave plate (HWP) and a polarization beam splitter (PBS), which pumps a temperature-stabilized periodically poled KTP (PPKTP) crystal placed in a polarization Sagnac interferometer. The interferometer consists of a dual-wavelength PBS and HWP (D-HWP) for 810 nm and 405 nm wavelengths. A plano-convex lens (focal length 150 mm) focuses the pump at the crystal center and collimates the photon pairs generated via spontaneous parametric down-conversion (SPDC). In the spatial division multiplexing section, the collimated SPDC emission ring, separated from the pump by a dichroic mirror (DCM), is divided into two halves using a prism mirror (PM), with each half further segmented into three subsections by D-shaped mirrors (DMs). Six diametrically opposite sections form three entangled photon sources, I, II, and III, with section pairs (1, 1'), (2, 2'), and (3, 3') as shown in the source division table. These subsections are multiplexed using three 50:50 beam splitters (BS), while mirrors (M) in the experiment direct the beams through the optical paths. The multiplexed outputs are transmitted via the communication channel to six users. Each user’s receiver consists of a projection and detection module comprising an HWP, quarter-wave plate (QWP), PBS, an interference filter (IF; bandwidth of 3 nm, centered at 810 nm), a fibre coupler, and a single-mode fibre (SMF) connected to a single-photon counting module (SPCM). The SPCM outputs are processed by a time-to-digital converter (TDC) and a computer. The user allocation table lists the allocated subsections for each user. The quantum correlation graph illustrates the realized quantum links within the network, which can be extrapolated to form a potential global quantum network based on satellite technology.}
    \label{Figure1}
\end{figure}

The schematic of the experimental implementation, shown in Fig. \ref{Figure1}, comprises three main sections: the central unit, the free-space channel, and the user stations. The central unit comprises the entangled photon source, along with spatial division and multiplexing modules. The free-space section serves as the communication channel, while each user incorporates projection and detection systems for photon measurement. The entangled photon source is based on the type-0, quasi-phase-matched, degenerate, non-collinear spontaneous parametric down-conversion (SPDC) process, producing an annular ring intensity distribution where temporally and spatially correlated photon pairs at 810 nm lie at diametrically opposite points. The resulting quantum state is the Bell state, $\ket{\phi^-} = (\ket{HH} - \ket{VV})/\sqrt{2}$. To generate multiple entangled-photon sources from a single SPDC emission ring, we segmented the ring into six spatial sections using a prism mirror (PM) and D-shaped mirrors (DM). This configuration yielded three pairs of diametrically opposed sections, each carrying correlated pair photons at random. The spatially separated sources corresponding to these diametric pairs are illustrated within Fig. \ref{Figure1} physical representation graph. The sections are then multiplexed using beam splitters and distributed through a one-meter free-space communication channel to different users in the laboratory. The assignment of the sections among different users is listed in user allocation table in Fig. \ref{Figure1}.

Each of the six users (Alice (A), Bob (B), Charlie (C), Dave (D), Evan (E), and Feng (F)) is equipped with a polarization projection system (waveplates and a polarizing beam splitter), a single-photon counting module, and a time-to-digital converter for detection and analysis. Alice \& Bob (AB), Charlie \& Dave (CD), and Evan \& Feng (EF) are connected through the two outports of the same beam splitter and, hence, do not have direct communication links between themselves. Nevertheless, each user is directly connected to four others, forming a twelve-channel current quantum network, as depicted in Fig. \ref{Figure1} quantum correlation graph. For future quantum networks, the source and multiplexing system integrated with a dedicated telescope for each channel could be deployed on a drone or satellite to enable long-distance free-space QKD links. For further details, refer to the Materials and Methods section.
\section{Results and discussions}

\begin{figure}[H]
    \centering
    \includegraphics[width=\linewidth]{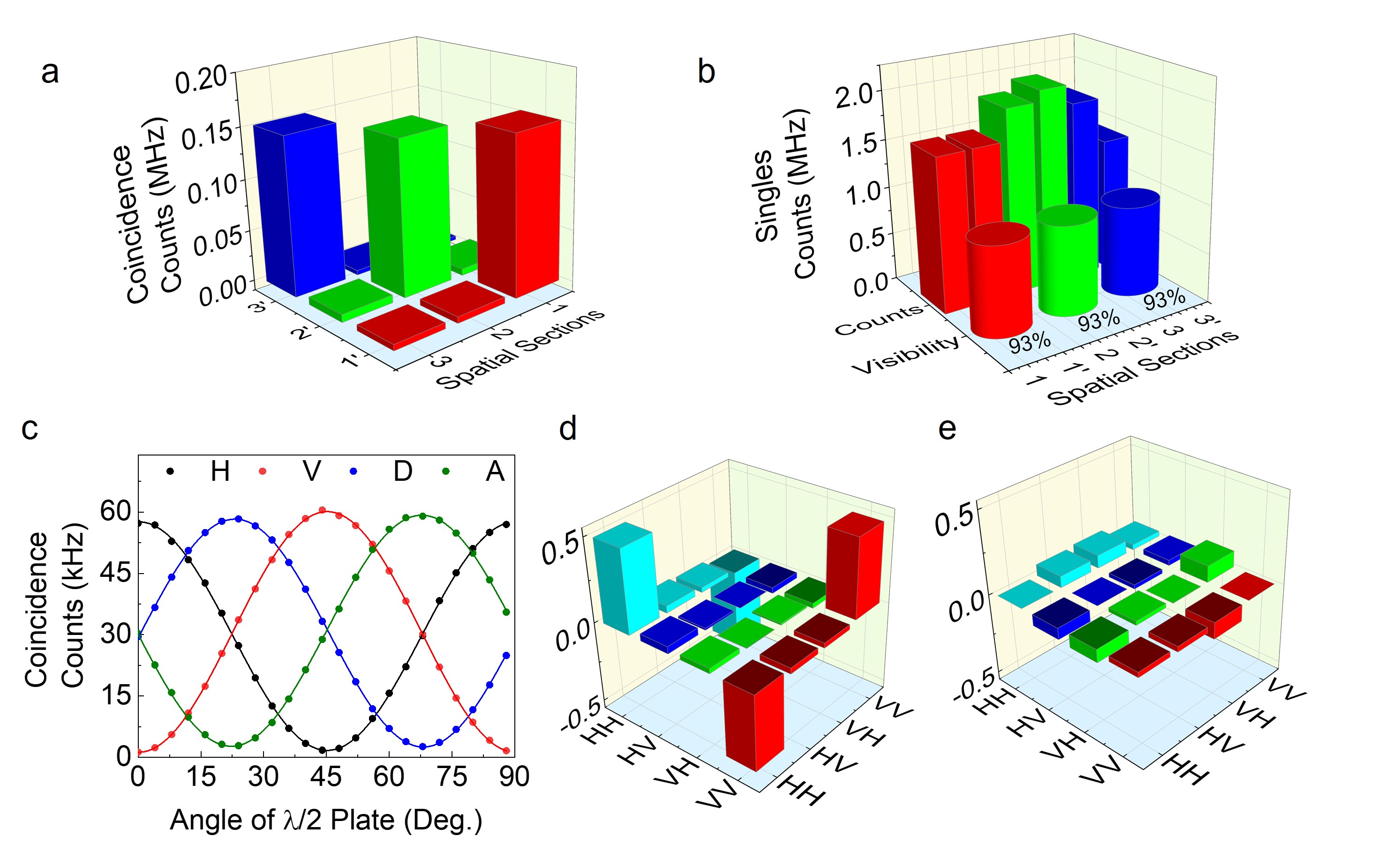}
    \caption{\textbf{Characterization of spatially separated entangled photon sources derived from a single SPDC source.} (a) Coincidence counts between different subsections, showing strong correlations between diametrically opposite sections carrying pair photons generated via the SPDC process. (b) Singles counts of different subsections, along with the coincidence visibility between diametrically opposite sections, 1 \& 1', 2 \& 2', and 3 \& 3'. (c) Quantum interference of the spatially separated entangled photon sources (here, we used source II comprises of subsections 2 \& 2') measured in the horizontal (H, black dots), vertical (V, red dots), diagonal (D, blue dots), and anti-diagonal (A, green dots) polarization bases. (d,e) Graphical representation of the absolute values of the real (d) and imaginary (e) parts of the reconstructed density matrix of the polarization-entangled Bell state $\ket{\phi^-}$.}
    \label{Figure2}
\end{figure}

We first characterized the entangled photon sources, with the results summarized in Fig. \ref{Figure2}. By pumping the nonlinear crystal with 1 mW of power, we tuned the phase-matching temperature to 39.5$^{\circ}$C to increase the SPDC ring diameter, thereby facilitating easier spatial sectioning. While a larger ring diameter by lowering the crystal temperature away from the degenerate phase-matching temperature enables finer division of the annular emission ring and increases the number of entangled photon sources derivable from a single setup, it also spreads the photon pairs over a larger area, reducing the collected photon flux and thus lowering the achievable bit rate. However, an increase in pump power can mitigate the issue of the lower bit rate. The measured coincidence counts (without polarization projection) between different subsections, as shown in Fig. \ref{Figure2}(a), confirm strong correlations between diametrically opposite sections, with a coincidence rate of $0.15 \times 10^6$ s$^{-1}$, and negligible ($6 \times 10^3$ s$^{-1}$) or no coincidences between non-opposite sections. The single-photon detection rates (Fig. \ref{Figure2}(b)) from six spatial subsections vary between $1.3 \times 10^6$ and $2.0 \times 10^6$ s$^{-1}$, corresponding to coupling efficiencies above 10$\%$. The variation in single counts among the subsections may be attributed to imperfect spatial sectioning \ayan{and collection} of the SPDC ring, likely caused by mechanical constraints in the experimental setup. Although this variation does not affect the present study, it could be minimized across all sections by employing an advanced spatial sectioning \ayan{and collection} approach using a microlens array coupled to fiber bundles. 

The polarization correlation visibility (for example between subsections \ayan{2 and 2' shown as violet color diamonds}) across all bases for the diametrically opposite sections exceeds 93\%, as illustrated in Fig. \ref{Figure2}(c), which shows the polarization-dependent coincidence counts for one representative entangled photon pair in the horizontal (H), vertical (V), diagonal (D), and anti-diagonal (A) bases. The measured Bell parameter is $S = 2.63$, clearly violating the classical limit of 2. We observe similar performance parameters while considering the pair subsections, 1 \& 1', and 3 \& 3'. Further, quantum state tomography performed by reconstructing both the real (Fig. \ref{Figure2}(d)) and imaginary (Fig. \ref{Figure2}(e)) components of the two-photon density matrix confirms that the generated state corresponds to the maximally entangled Bell state $\ket{\phi^-} = (\ket{HH} - \ket{VV})/\sqrt{2}$, with a state fidelity of 98\%. These observations collectively confirm the reliable and high-quality generation of three independent entangled photon sources derived from a single experimental resource. Using the difference in the number of reflections (even or odd) in the two arms of a single source, we characterized the quantum state as $\ket{\phi^-}$ and $\ket{\phi^+}$, with the same entangled parameters, as expected.

\begin{figure}[H]
    \centering
    \includegraphics[width=\linewidth]{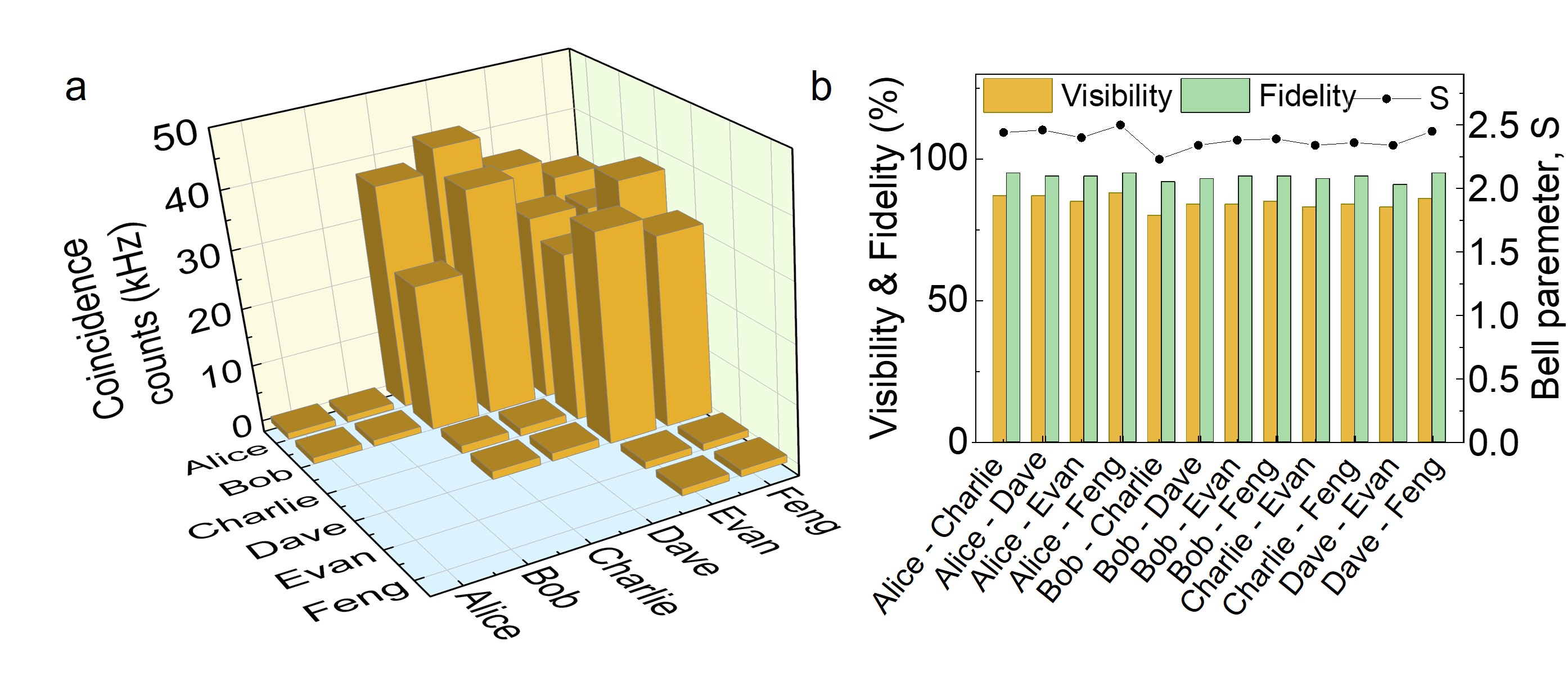}
    \caption{\textbf{Performance characteristics of quantum links within the network.} (a) Coincidence counts for each link connecting two users in the 12-link quantum network. (b) Measured quantum parameters for each link, including entanglement visibility, Bell parameter ($S$), and quantum state fidelity. The network supports both $\ket{\phi^+}$ and $\ket{\phi^-}$ Bell states, with each state distributed across six of the twelve links owing to the network's topology. }
    \label{Figure3}
\end{figure}

Having characterized the performance of the three spatially separated entangled photon sources derived from a single setup, we investigated the properties of a quantum network established through their multiplexing using 50:50 beamsplitters. The results are presented in Fig. \ref{Figure3}. As shown in Fig. \ref{Figure3}(a), each user shares strong correlations with four others, with coincidence rates ranging from $0.2 \times 10^5$ s$^{-1}$ to $0.4 \times 10^5$ s$^{-1}$, confirming the formation of a twelve-channel quantum network among six users. 
The observed variation in coincidence counts among users may be attributed to imperfect spatial sectioning using the available components in the laboratory and collection of the SPDC ring, implemented to accommodate mechanical constraints of the experimental setup, as well as deviations of the beam splitters from ideal 50:50 splitting ratios. On the other hand, the users connected via the same output port of a beam splitter such as Alice \& Bob, Charlie \& Dave, and Evan \& Feng exhibit only minimal coincidence counts ($1.1 \times 10^3$~s$^{-1}$ to $1.4 \times 10^3$~s$^{-1}$), attributed to background noise, indicating no direct communication link between them. To further assess channel integrity, we introduced digital delays up to 20 ns between pair photons across different user combinations. The resulting coincidence counts remained within a similar noise-limited range, stemming from dark counts, accidental events, and optical imperfections in the system. It is to be noted that all the results presented in this article are based on raw, unprocessed data. While standard correction techniques accounting for detector inefficiencies and background subtraction can improve apparent performance, they may obscure the true practicality of the scheme. To retain relevance for real-world applications, no such corrections were applied in this analysis.

Furthermore, we measured the entanglement visibility, Bell parameter, and quantum state fidelity for each link Alice-Charlie (AC), Alice-Dave (AD), Alice-Evan (AE), Alice-Feng (AF), Bob-Charlie (BC), Bob-Dave (BD), Bob-Evan (BE), Bob-Feng (BF), Charlie-Evan (CE), Charlie-Feng (CF), Dave-Evan (DE), and Dave-Feng (DF), connecting any two users in the network (see Fig. \ref{Figure1}). As shown by the golden bars in Fig. \ref{Figure3}(b), the interference visibility, averaged over all polarization bases for each user pair in the network, ranges from 80\% to 88\%. This is slightly lower than the visibility observed from the individual sources, primarily due to imperfections introduced by additional optical elements, including the beam splitter in the network. Nevertheless, the measured visibilities remain well above the 71\% threshold required to confirm entanglement \cite{Jabir2017}, thereby verifying entanglement across all users in the network.

We also performed quantum state tomography shared between two user pairs in the network and found that due to the use of different numbers of reflections in the network paths the links, AC, AF, BC, BF, CE, and DE and AD, AE, BD, BE, CF, and DF connect two users through $\ket{\phi^+}$ and $\ket{\phi^-}$, respectively. The measured fidelity of the shared quantum states varies in the range from 91\% to 95\% (see green bars in Fig. \ref{Figure3}(b)). The corresponding Bell parameters $S$ also exceeded the classical limit of $S = 2$, varying in the range of $S = \text{2.2} - \text{2.5}$, thus confirming the presence of quantum correlations. Using the coincidence counts between each pair of users connected through the twelve-channel network in the HV and DA bases, and applying binary Shannon entropy, we estimated the secure key rate between every pair (see Eq. \ref{Eq.1} in the Methods section), yielding a total secure key rate of 76 kbps across the network. Similarly, the total sifted key rate, calculated from the coincidence counts in the HV and DA bases, is found to be 407 kbps. These numbers will be high for the detection using SNSPD detectors with quantum efficiency $>$95$\%$. As explained previously, the use of standard corrections will improve the values of the parameters close to ideal. \gks{However, these results demonstrate the successful realization of a twelve-channel entanglement-based quantum network, enabling simultaneous quantum key distribution among six users at high bit rates. With improved collection strategies, use of SNSPD detectors, and an increased number of spatial sections while considering a larger diameter SPDC ring generated at high pump power, this spatial-division multiplexing technique holds promise for scaling up multi-user quantum communication networks beyond what is currently feasible using WDM schemes. }
\begin{figure}[H]
    \centering
    \includegraphics[width=\linewidth]{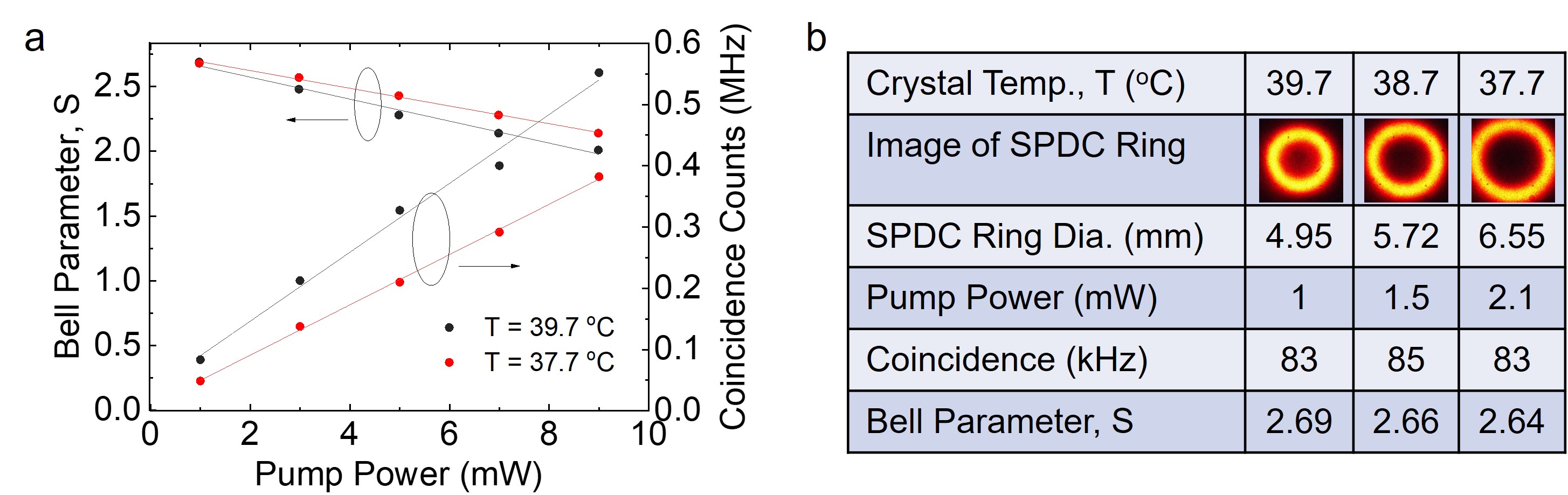}
    \caption{\textbf{Performance parameters of the entangled-photon source for scalable quantum networking.} (a) Variation of the Bell parameter $S$ and coincidence counts as a function of pump power for two spontaneous parametric down-conversion (SPDC) emission-ring diameters, corresponding to crystal temperatures of $T = 37.7^{\circ}$C and $T = 39.7^{\circ}$C. (b) Tabulated SPDC source characteristics at the two crystal temperatures, showing that an increased emission-ring diameter allows finer spatial sectioning. This capability supports network-topology scaling to accommodate more users while preserving the entanglement quality of the source. }
    \label{Figure4}
\end{figure}

To assess the scalability of our approach, we performed a separate experiment evaluating the performance of the entangled photon source in terms of the Bell parameter $S$ and coincidence counts as a function of pump power at two phase-matching temperatures, $T$. The results are shown in Fig. \ref{Figure4}. As evident from Fig. \ref{Figure4}(a), at T = 39.7$^o$C, the rate of decrease in $S$ with pump power is measured to be 0.085 mW$^{-1}$, while the coincidence count rate increased at 0.056 MHz$/$mW. Reduction of crystal temperature to T = 37.7$^o$C, away from the degenerate, collinear phase-matching condition,  enlarged the SPDC emission ring, decreased the coincidence count slope to 0.041 MHz$/$mW, and hence the attainable bit rate. However, the smaller degradation rate of $S$ = 0.068 mW$^{-1}$ suggests that higher pump powers can be employed to recover the bit rate without compromising quantum correlations. This behaviour can be attributed to the reduced photon density across the enlarged ring \cite{Jabir2017}, which, in addition to lowering the bit rate, lowers the probability of multi-pair generation at the same spatial location, a primary source of correlation degradation. Consequently, the source preserves comparable quantum performance and bit rate even at elevated pump powers, demonstrating robustness to variations in ring size. To further quantify this, we have tabulated key performance parameters between two users for different ring sizes. As shown in Fig. \ref{Figure4}(b), lowering the crystal temperature from 39.7$^o$C to 37.7$^o$C increases the annular ring diameter from 4.95 mm to 6.55 mm, while still achieving comparable coincidence rates ($\sim$ 83 kHz) and Bell parameters (S $>$ 2.6) with increased pump power (from 1 mW to 2.1 mW). These results clearly demonstrate the scalability of our entangled photon source for quantum networks with a larger number of users. Again, given the higher bit rate achieved in the network, each user can be considered as a node, enabling the expansion of the network by employing the WDM schemes \cite{wengerowsky18, joshi2020} or a one-to-multi-port beam splitter \cite{Qi21} to connect additional users to the network. This approach facilitates the realization of a large-scale quantum network.

\section{Conclusion}

We have demonstrated a scalable quantum network architecture based on spatial division and beam-splitter-based multiplexing of an SPDC ring, enabling simultaneous distribution of polarization-entangled photon pairs among multiple users. In contrast to conventional schemes relying on WDM or multi-core fiber arrays, our approach eliminates the need for narrowband filters and complex spatial coupling optics, thereby preserving high photon collection efficiency while minimizing inter-channel crosstalk.

The network supports dynamic reconfiguration where the users can be added or removed without altering the core optical setup, making it highly adaptable for real-world deployment. Using polarization measurements in mutually unbiased bases (HV and DA), we estimated a total sifted key rate of 407 kbps and a secure key rate of 76 kbps across all user pairs, demonstrating the viability of the architecture for high-throughput quantum key distribution.

Scalability can be further enhanced by increasing the diameter of the SPDC ring and pump power, enabling more communication channels without compromising the bit rate. Moreover, the integration of the current scheme with WDM devices or one-to-many beam splitters offers a route toward building even larger quantum networks. Our results establish a versatile and efficient platform for a next-generation global quantum network for multi-user quantum communication, overcoming key limitations of WDM-based approaches in terms of spectral channel constraints and inter-channel crosstalk.

\section*{ACKNOWLEDGMENTS}
A. K. N. and G. K. S. acknowledge the support of the Department of Space, Govt. of India. A. K. N. acknowledges funding support for Chanakya - PhD fellowship from the National Mission on Interdisciplinary Cyber-Physical Systems of the Department of Science and Technology, Govt. of India through the I-HUB Quantum Technology Foundation.  G. K. S. acknowledges the support of the Department of Science and Technology, Govt. of India, through the Technology Development Program (Project DST/TDT/TDP-03/2022). 

\section*{AUTHOR DECLARATIONS}
\subsection*{Conflict of Interest}
The authors have no conflicts to disclose.
\subsection*{Author Contributions}
A. K. N. developed the experimental setup, performed measurements, data analysis, numerical simulation, and data interpretation. G.K.S. participated in data analysis and data interpretation. G. K. S. developed the ideas and led the project. All authors participated in the discussion and contributed to the manuscript writing.

\section*{DATA AVAILABILITY}
The data that support the findings of this study are available from the corresponding author upon reasonable request.

\section{Materials and Methods}
\subsection{Polarization entangled photon source}
The schematic of the experimental setup is shown in Fig.\ref{Figure1}. We used a continuous-wave, single-frequency diode laser that delivers an output power of 21 mW at a central wavelength of 405 nm with an approximate linewidth of 20 MHz as the pump source for the experiment. To improve the spatial mode quality of the laser, we couple its output into a single-mode fiber (SMF) and subsequently collimate it using a fiber collimator. The laser power to the experiment is controlled using a combination of a half-wave plate (HWP) and a polarizing beam splitter (PBS) cube. Another HWP is used to rotate the horizontal polarized pump beam to a diagonal polarization state, ensuring equal power distribution between the two output ports of a dual-wavelength PBS (D-PBS), and hence equal power to the clockwise (CW) and counterclockwise (CCW) propagating pump beams inside a polarization Sagnac interferometer. The Sagnac interferometer is comprised of D-PBS, optimized for both 405 nm and 810 nm, and two plane mirrors (M). A lens (L) of focal length $f = 150$ mm focuses the pump beam from both directions to a spot diameter of $\sim$40 $\mu$m at the center of the nonlinear crystal, which is symmetrically placed between two mirrors, M, relative to the D-PBS. A 20 mm-long periodically poled potassium titanyl phosphate (PPKTP) crystal with a grating period $\Lambda = 3.425$ $\mu$m and an aperture of $1 \times 2$~mm$^2$ is used to generate non-collinear, degenerate, type-0 (e $\rightarrow$ e + e), quasi-phase-matched pair photons at 810 nm through the spontaneous parametric down-conversion (SPDC) process. The crystal is housed in a temperature-controlled oven, which can be varied up to 200$^\circ$C with a stability of $\pm$0.1$^\circ$C to maintain optimal phase matching.

A dual-wavelength half-wave plate (D-HWP), optimized for 405 nm and 810 nm and oriented at 45$^\circ$ to the vertical axis, is placed inside the Sagnac interferometer to transform the horizontal polarization into vertical and vice versa. As described in Refs.\cite{Jabir2017, Singh2023}, the SPDC photons generated from the CW and CCW beams exhibit annular ring distributions, with pair photons lying diametrically opposite points on the ring, are recombined at the D-PBS and collimated using lens L. The dichroic mirror (DCM), placed between lens L and the HWP, separates the SPDC photons from the pump. The resulting entangled quantum state of the pair photons is given by $\ket{\phi^\pm} = (\ket{HH} - \ket{VV})/\sqrt{2}$, representing a Bell state. The SPDC ring is captured using an electron-multiplying charge-coupled device (EMCCD) camera (Andor iXon Ultra 897) with an interference filter centered at 810 nm and a bandwidth of approximately $\sim$3.2 nm, as shown in the inset of Fig. \ref{Figure1}.

\subsection{Spatial division and multiplexing}
Since entangled photon pairs lie on diametrically opposite points of the SPDC ring, we segmented the ring to create three independent entangled photon sources. First, we used a gold-coated prism mirror (PM) to divide the annular SPDC ring into two equal halves. We then further subdivided each half into three angular sections using four D-shaped mirrors (DM), creating two sets of spatial segments, (1, 2, 3) and (1', 2', 3'). These segments are arranged (see Fig. \ref{Figure1} physical representation graph such that each pair of diametrically opposite sections, (1, 1'), (2, 2'), and (3, 3') contains correlated pair photons forming three spatially separated entangled photon sources, I, II, and III, respectively, from a single set of resources: the pump laser, the nonlinear crystal, and the Sagnac interferometer. 

We further multiplexed the six subsections of the three entangled photon sources and assigned communication channels using three nearly 50:50 beam splitters (BS), as illustrated in Fig. \ref{Figure1} experimental setup. Subsections 1 and 2, originating from entangled photon sources I and II, respectively, are combined through two input ports of a beam splitter, and the two output ports are routed to users Alice (A) and Bob (B). As a result, photons reaching Alice and Bob originate from sources I and II with equal probability. However, due to the probabilistic nature of beam splitter outputs, Alice and Bob do not receive photon pairs simultaneously, and therefore, no direct quantum communication is possible between them.

Similarly, subsections (3, 1') from entangled photon sources III and I, and (2', 3') from sources II and III, are each merged at beam splitters, with their outputs directed to users Charlie (C), Dave (D), and Evan (E), Feng (F), respectively. Like Alice and Bob, these user pairs (Charlie, Dave) and (Evan, Feng) are connected by the same beam splitter, and hence cannot perform quantum communication between themselves.

However, this multiplexed configuration enables each of the six users to share entanglement-based communication with four other users via three distinct entangled photon sources, forming a twelve-channel quantum network, as shown in Fig. \ref{Figure1} quantum correlation graph. 

\subsection{Projection and detection for each user}
All six quantum channels guided using mirrors, M in the experiment, are equidistant (although a necessary condition) in free space from the BS to the user. To measure the entanglement among the parties, each user has a polarization projection system comprised of either HWP, PBS, or HWP, QWP, and PBS, and the detection systems comprised of an interference filter (bandwidth $\sim$3.2 nm), mounted fiber coupler, single-mode fiber (SMF), and single-photon counting modules, SPCMs (SPCM-AQRH-14-FC). All SPCMs are connected to a single time-to-digital converter (TDC) (Swabian, Time Tagger Ultra) and a computer, for photon counting, time stamping, recording, and experimental data analysis. 

\subsection{Estimation of secure key rate}
The secure key rate between any two users in the network can be estimated based on the formula presented in Ref. \cite{ortega2021}

\begin{align}
\label{Eq.1}
R_{SK} &= \frac{1}{2} C_{DA} \left[ 1 - (1 + m) H\left( \frac{1 - V_{DA}}{2} \right) \right] \notag \\
&\quad + \frac{1}{2} C_{HV} \left[ 1 - (1 + m) H\left( \frac{1 - V_{HV}}{2} \right) \right]
\end{align}

Here, $C_{HV}$ and $C_{DA}$ are the total numbers of coincidence counts in the DA and HV bases, H is the binary Shannon entropy, and we used a bidirectional error correction of m = 1.1.

\subsection{Scaling of the network}
The presented network architecture is highly scalable, allowing users to be added or removed without altering the core design. In contrast, prior networks based on wavelength division multiplexing (WDM) require additional narrow-bandpass filters symmetric to the degenerate wavelength of the entangled photon source, defined by ITU frequency channels. However, such approaches reduce the bit rate between users and increase crosstalk. Moreover, scaling the WDM-based network by increasing the spectral bandwidth of the entangled photon source demands a shorter nonlinear crystal length, which decreases the source brightness. In addition, an increase in the bandwidth of the source requires phase correction of the source arising from the dispersion of the optical elements present in the experiment and the length of the fiber link.

Spatial multiplexing using multi-core fiber also poses challenges: a larger number of fiber cores reduces the collection efficiency of photons, thereby lowering the overall bit rate. These limitations are overcome in the present network. Scaling can be achieved by increasing the diameter of the SPDC ring; however, for a given input power and figure-of-merit of the nonlinear crystal, photon pair density over a larger annular area decreases collection efficiency and channel bit rate. This can be mitigated by increasing the pump power (see Fig. \ref{Figure4}), thereby generating photon pairs at a higher rate to compensate for the spatial spread.

Furthermore, considering each user as a node, additional scalability can be achieved by integrating WDM devices and/or one-to-many beam splitters. Employing faster detection systems and reducing the coincidence window allows higher source brightness and mitigates interlink crosstalk. Importantly, simultaneous communication between network nodes is not hindered by the absence of a node or its users, except that users connected through that node become inaccessible for communication.

\bibliography{sn-bibliography}

\end{document}